\documentclass[twocolumn,aps,prl,showpacs,groupedaddress]{revtex4-1}

\usepackage{amsmath}
\usepackage{dcolumn}
\usepackage{epsfig}
\usepackage{graphicx}
\usepackage{latexsym}

\usepackage{epstopdf}

\begin{document}

\title{Bragg Scattering as a Probe of Atomic Wavefunctions and Quantum Phase Transitions in Optical Lattices}

\author{Hirokazu Miyake}
\author{Georgios A. Siviloglou}
\author{Graciana Puentes}
\author{David E. Pritchard}
\author{Wolfgang Ketterle}
\author{David M. Weld}
\affiliation{MIT-Harvard Center for Ultracold Atoms, Research Laboratory of Electronics, and Department of Physics, Massachusetts Institute of Technology, Cambridge, Massachusetts 02139, USA}

\begin{abstract}
We have observed Bragg scattering of photons from quantum degenerate $^{87}$Rb atoms in a three-dimensional optical lattice.  Bragg scattered light directly probes the microscopic crystal structure and atomic wavefunction whose position and momentum width is Heisenberg-limited.  The spatial coherence of the wavefunction leads to revivals in the Bragg scattered light due to the atomic Talbot effect.  The decay of revivals across the superfluid to Mott insulator transition indicates the loss of superfluid coherence.
\end{abstract}

\pacs{05.30.Rt, 37.10.Jk, 61.05.-a, 67.85.-d}

\maketitle 

Ultracold atomic gases are an ideal system in which to study many-body phenomena because of the relative ease with which parameters in the model Hamiltonian can be tuned across a wide range~\cite{lewenstein07,bloch08}.  Such studies have resulted in a better understanding of various phase transitions such as the Berezinskii-Kosterlitz-Thouless transition in two dimensional systems~\cite{hadzibabic06}, the BEC-BCS crossover of interacting fermions~\cite{zwierlein05} and the superfluid to Mott insulator transition in a three-dimensional lattice~\cite{greiner02}.  One major goal of this field is to realize spin phases such as antiferromagnetic states to explore quantum magnetism and its interplay with high-temperature superconductivity~\cite{lee06}.

Concurrently, there are numerous efforts to develop techniques to probe and understand the atomic ensemble once a new type of ordering is achieved.  One recent development is the realization of single-site resolution of atoms in two-dimensional optical lattices~\cite{bakr09,sherson10}.  An alternative method to measure \textit{in situ} spatial ordering is the technique of Bragg scattering, often used in a condensed matter context to determine crystal structure~\cite{ashcroft76}.  In particular, Bragg scattering with neutrons led to the discovery of antiferromagnetism in the solid state~\cite{shull49} and with x-rays led to the discovery of the double helix structure of DNA~\cite{watson53}.

Bragg scattering relies on the interference of waves scattered coherently from an ensemble of particles.  In particular when atoms are arranged in a periodic pattern in three spatial dimensions, there are certain angles of the incoming and reflected beams where scattering is dramatically enhanced compared to other angles.  This has allowed crystallographers to use x-rays to determine the properties of crystals such as lattice geometry at the \r{a}ngstrom scale.  We have applied this technique to ultracold atoms in a three-dimensional optical lattice by scattering near-infrared light where the atoms are spaced almost $10^4$ times farther apart than those in typical condensed matter samples.

\begin{figure}
\begin{center}
\includegraphics[width=1.0\columnwidth]{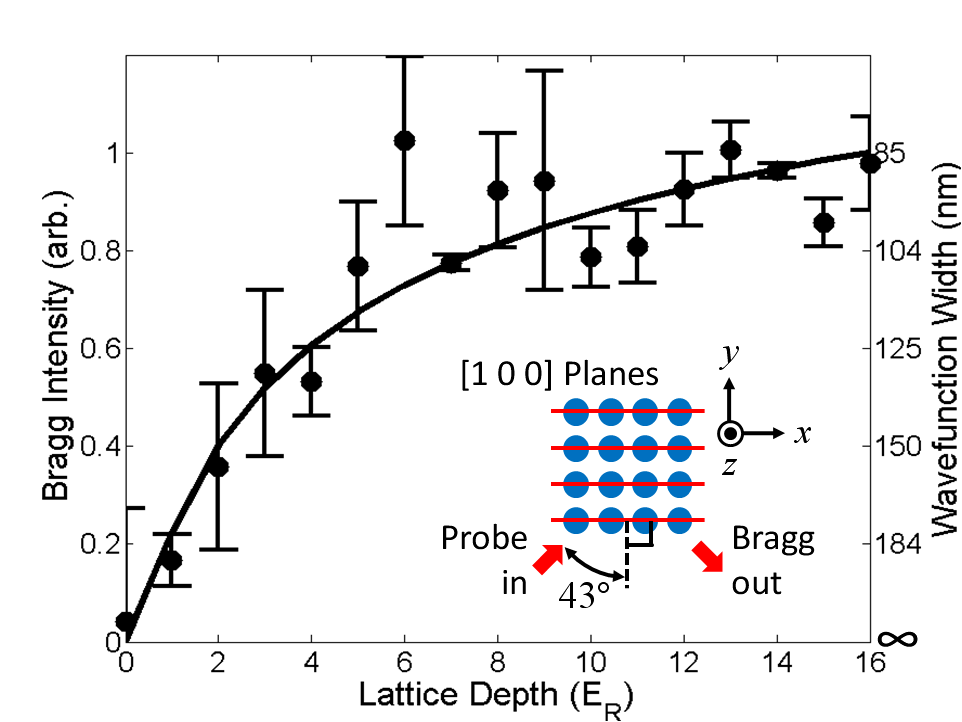}
\caption{Bragg scattering as a probe of the spatial wavefunction width.  Bragg scattered intensity vs. lattice depth in units of the recoil energy.  The right axis gives the corresponding root-mean-square width of the wavefunction squared.  The solid line is a no-free-parameter curve given by the Debye-Waller factor.  Error bars are statistical.  Inset is a schematic of the setup for Bragg scattering.
\label{fig:countvlatt}}
\end{center}
\end{figure}

Pioneering works on Bragg scattering from cold atoms in optical lattices were done by the H\"{a}nsch and Phillips groups using laser cooled atoms~\cite{weidemuller95}.  These lattices were sparsely populated and the atoms occupied several bands.
In this Letter, we have used Bragg scattering to study bosonic atoms cooled to quantum degeneracy and placed in a three dimensional cubic lattice where the atoms occupy the lowest band.
In particular, we have measured directly the Heisenberg-limited width of both position and momentum of the single ground state atomic wavefunction in an optical lattice.  Furthermore, there is a revival of Bragg scattered light some time after releasing the atoms from the optical lattice, analogous to the optical Talbot effect.  This signal gives a direct measure of the coherence of the superfluid component in the lattice.

The experimental apparatus has been described in detail elsewhere~\cite{streed06}.  Briefly, laser cooling and evaporative cooling are used to achieve quantum degeneracy of $^{87}$Rb atoms in the $|F=2,m_F=-2 \rangle$ state which are loaded into a crossed dipole trap whose trap frequencies range between 30 and 160 Hz.  Once quantum degeneracy is achieved, the optical lattices generated by a single 1064 nm fiber laser are adiabatically ramped on.  The lattices are calibrated by applying a 12.5 $\mu$s pulse and measuring the Kapitza-Dirac diffraction of the atoms and comparing this to theory. The system typically contains about $10^5$ atoms, which leads to up to 5 atoms per lattice site.  The Bragg reflected light is detected on a CCD camera which images along a direction which satisifies the Bragg condition.

The Bragg scattering condition for a three-dimensional cubic lattice is given by $2 d \sin \theta = n \lambda_p$
where $d$ is the spacing between lattice planes, $\theta$ is the angle between the incoming beam and the lattice planes, $n$ is any positive integer and $\lambda_p$ is the wavelength of the probe beam.  For our experiment $\lambda_p$ is 780 nm and $d$ is 532 nm.  With these conditions the only allowed angle $\theta$ is $47^{\circ}$ where $n$ is one and corresponds to Bragg reflection off the [1 0 0] plane or any equivalent plane.  A schematic of the probe beam with respect to the atoms is shown in the inset of Fig.~\ref{fig:countvlatt}.  Since the full angular width of the Bragg scattered light is small (measured to be $4.1 \pm 0.4^{\circ}$), a precise alignment of the incident beam had to be performed at the 1/min repetition rate of the experiment.  In contrast, in one and two dimensions, diffractive light scattering occurs at any angle of incidence, as recently shown with atoms in a two-dimensional optical lattice~\cite{weitenberg11}.

\begin{figure}
\begin{center}
\includegraphics[width=1.0\columnwidth]{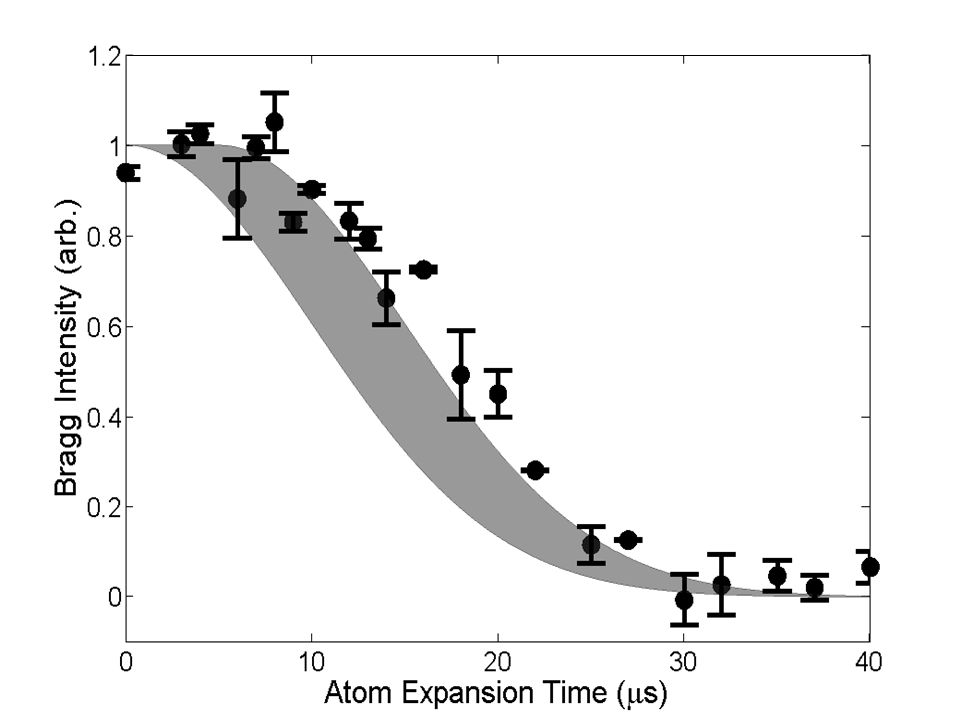}
\caption{Bragg scattering as a probe of the momentum wavefunction width.  Bragg scattered intensity vs. the free-expansion time of the atoms after rapidly turning off the lattices from $15E_{\rm R}$.  The decay in signal indicates a melting of the crystal structure, or in other words spreading of the atomic wavepacket with a given momentum uncertainty.  The gray area is a no-free-parameter curve using the Debye-Waller factor taking into account the probe pulse duration of 5 $\mu$s.  Error bars are statistical.
\label{fig:wfspread}}
\end{center}
\end{figure}

The probe beam at the Bragg angle had a typical power of 0.3 mW and beam diameter of 300 $\mu$m, large enough to illuminate the entire atomic cloud, and was detuned from the $5^{2}S_{1/2}, F=2 \rightarrow 5^{2}P_{3/2},F'=3$ cycling transition of $^{87}$Rb by a few natural linewidths, where the natural linewidth $\Gamma$ is 6 MHz.  The detuning needs to be sufficient so that the light traverses the entire atom cloud.  Although the absolute signal varies with the detuning, the conclusions reached in this Letter appear not to depend strongly on the amplitude or sign of the detuning.  The probe beam was applied for a few microseconds, enough to obtain a signal but short enough to avoid heating effects.

As the lattice is increased the Bragg reflected signal increases as expected from the crystal ordering and tighter localization of the atoms (Fig.~\ref{fig:countvlatt}).
%where the data were taken at a detuning of $-5 \Gamma$.
For this data, the lattice in the $z$ direction, which is also the direction of gravity, was ramped to 15$E_{\rm R}$, where $E_{\rm R}=h^2/(2m \lambda^2_L)$ is the recoil energy, $h$ is the Planck constant, $m$ is the mass of $^{87}$Rb and $\lambda_L = 1064$ nm is the lattice laser wavelength.  Simultaneously the lattices in the horizontal directions were ramped to various lattice depths.  Note that for our geometry the Bragg reflected intensity is insensitive to the lattice depth in the $z$ direction because Bragg scattering occurs only in the horizontal $xy$ plane, as we discuss later.

The scattering of light by a collection of atoms can be modeled in the following way.  In the limit of low probe intensity and low optical density of the cloud, the Born approximation can be used.  Then the scattering cross-section $d \sigma/d \Omega$ can be written as a product of one-dimensional Debye-Waller factors where
$d \sigma/d \Omega \propto \prod_i \exp(-(\Delta r_i)^2 K_i^2/2)$,
where $i=x,y,z$ is the index for the three dimensions, $K_i$ is the magnitude of the $i$-th component of $\textbf{K}$ where $\textbf{K} = \textbf{k}_{\rm in} - \textbf{k}_{\rm out}$ is the difference between the incoming probe beam and Bragg scattered wavevectors, and $\Delta r_i = \sqrt{\hbar /(m \omega_i)}$ is the harmonic oscillator width where $\hbar$ is the reduced Planck constant and $\omega_i$ is the trap frequency in the $i$-th direction, which depends on the optical lattice depth.  The atoms are approximated as gaussian wavefunctions, which is a good approximation for sufficiently large lattice depths.  However, we find that even for low lattice depths where significant superfluid components are expected, this approximation describes the Bragg scattered signal well.

For our experimental conditions, $K_z = 0$ because Bragg scattering occurs in the horizontal $xy$ plane.  The lattice depths are controlled in a way such that they are the same in both the $x$ and $y$ directions which leads to $\Delta x \equiv \Delta r_x = \Delta r_y$.  This simplifies the scattering cross-section to $d \sigma/d \Omega \propto \exp(-(\Delta x)^2 K^2/2)$.  Thus Bragg scattering allows the study of the spatial extent of the atomic wavefunction.

The data can be compared to a no-free-parameter theoretical line, assuming non-interacting atoms.  The Bragg scattered intensity $B(N_L)$ as a function of lattice depth $N_L$ is proportional to $\exp(-\lambda^2_L K^2/(8 \pi^2 \sqrt{N_L}))$.  Both the harmonic oscillator width $\Delta x$, which depends on the mass of the rubidium atom and trap frequency, and change in wavevector $K$ are known.  This theoretical line and the data is shown in Fig.~\ref{fig:countvlatt} and we see good agreement.  Thus the Bragg scattered light as a function of lattice depth can be well-described by a model where the atoms are assumed to be non-interacting gaussian atomic wavefunctions whose spatial width is determined by the lattice depth.

Bragg scattering can also be used to probe the momentum width of the atomic wavefunctions.  This is done by measuring Bragg reflection as a function of the expansion time of the atoms between a rapid lattice turn off and Bragg probe.
In particular, the two horizontal lattices were turned off from $15E_{\rm R}$ to $0E_{\rm R}$ in less than 1 $\mu$s and the lattice in the $z$ direction was kept at $15E_{\rm R}$.  Turning off the lattices allows the atomic wavepackets in each individual lattice well to expand freely in those directions.  The data in Fig.~\ref{fig:wfspread} shows that the signal has decayed in 30 $\mu$s.  The time it takes to lose crystal structure is roughly the time it takes for the atoms to move over half of a lattice distance.

The Debye-Waller factor can be used to determine more precisely how the Bragg reflection should behave as a function of the expansion time with no free parameters, assuming gaussian atomic wavepackets.  This makes use of the well-known time-dependent behavior of a Heisenberg-limited gaussian wavepacket in free space which can be written as $(\Delta x(t))^2 = (\Delta x)^2 + (\Delta p)^2 t^2/m^2$ where $\Delta x$ and $\Delta p$ are the uncertainty of position and momentum respectively at the initial time and $t$ is the expansion time~\cite{schiff68}.  In a previous paper where the atoms were not quantum degenerate, the momentum distribution was assumed to be determined by temperature~\cite{weidemuller95}.  The results in this Letter show that the momentum uncertainty is Heisenberg-limited, meaning $(\Delta x)^2 (\Delta p)^2 = \hbar^2/4$, so the Bragg scattered light $B(t)$ as a function of expansion time decays as $\exp(-(\Delta p)^2 K^2 t^2/(2 m^2))$.  This curve is also shown in Fig.~\ref{fig:wfspread}.  The curve is broadened by taking into account the probe beam duration which was 5 $\mu$s and shows good agreement with data.
This analysis shows that by releasing the atoms from a lattice, we can directly probe the \textit{in situ} time evolution of the ground state atomic wavefunctions with spatial extent of tens of nanometers.  Furthermore, the atomic spatial and momentum widths are seen to be Heisenberg-limited.

\begin{figure}
\begin{center}
\includegraphics[width=1.0\columnwidth]{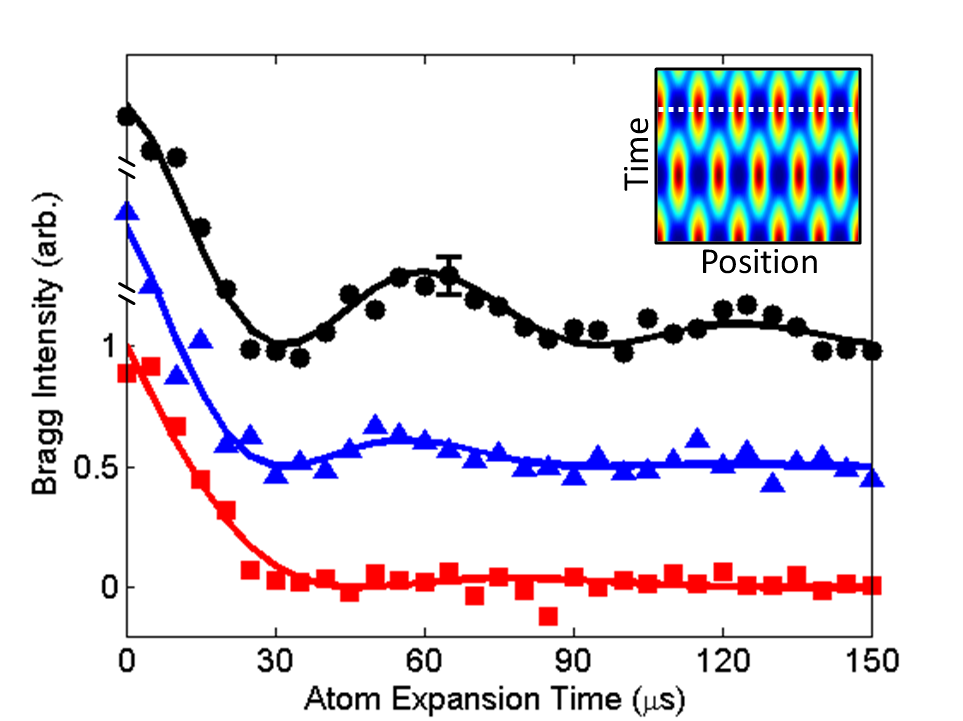}
\caption{Revivals of Bragg scattered light.  Bragg scattered intensity vs. the expansion time in three dimensions for three different initial lattice depths: $5E_{\rm R}$ (circles), $8E_{\rm R}$ (triangles), and $15 E_{\rm R}$ (squares).  Each data set is normalized to the Bragg intensity at the initial time.  Revivals at lower lattice depths indicate coherence of the atoms across multiple lattice sites.  The lines are phenomenological fits to exponentially decaying sine waves.  Different lattice depth data are offset for clarity and the error bar is represenative.  Inset is a solution to the one-dimensional Gross-Pitaevskii equation with experimental parameters and an initial state of a chain of gaussian wavefunctions, showing the revivals of the density distribution as a function of time.  The white dotted line represents the revival time.
\label{fig:revival_vs_time}}
\end{center}
\end{figure}

Coherent many-body non-equilibrium dynamics can also be studied with Bragg scattering after a sudden release of the atoms from the optical lattice in three dimensions.  At low lattice depths the Bragg scattered signal shows revivals as a function of expansion time because of the rephasing of the superfluid atomic wavefunctions that were originally confined in the lattices.  Therefore, these revival signals should give a measure of the superfluid order parameter.  Furthermore, the revivals are analogous to the optical Talbot effect, whose atomic version was observed previously for a thermal beam~\cite{chapman95} and for a quantum degenerate gas in the temporal domain~\cite{deng99}.  Here we observe the atomic Talbot effect for an interacting quantum degenerate gas in a three-dimensional optical lattice.  Revivals can be seen in Fig.~\ref{fig:revival_vs_time} and become less pronounced as the lattice depth increases.

In particular, the characteristic revival time is determined by $\tau = h/(\hbar^2 (2 k)^2/(2 m))$ and is 123 $\mu$s for our parameters, where $2k$ corresponds to the wavevector of the matter wave.  However, with Bragg scattering we expect the first revival at half that time of 61 $\mu$s, which is what we observe in Fig.~\ref{fig:revival_vs_time}.  This can be understood by realizing that the atomic wavepackets need only travel half the lattice spacing to constructively interfere with the wavepackets traveling in the opposite directions from the nearest neighbor sites.

\begin{figure}
\begin{center}
\includegraphics[width=1.00\columnwidth]{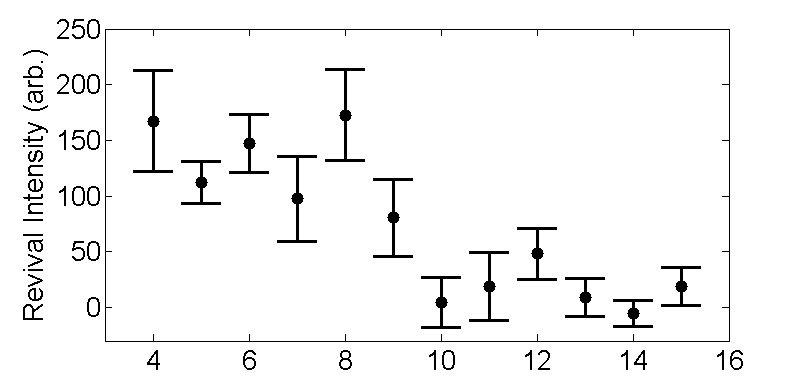}
\includegraphics[width=1.00\columnwidth]{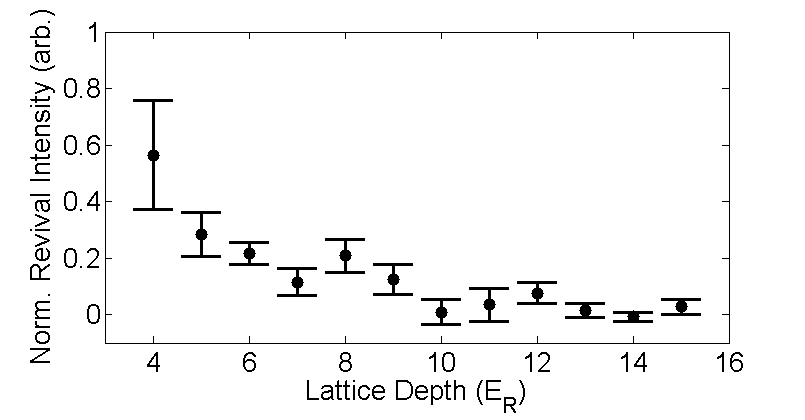}
\caption{Bragg scattering revival as a probe of superfluid coherence.  Top: Bragg scattered revival signal at 60 $\mu$s vs. lattice depth.  Bottom: The top plot normalized to the Bragg scattered signal without expansion.  We see a decrease in revival, and consequently superfluid coherence, as the lattice is increased.  Error bars are statistical.
\label{fig:revival_vs_latticedepth}}
\end{center}
\end{figure}

The superfluid coherence as a function of lattice depth can be studied by comparing the Bragg reflected signal when the atoms are in the optical lattice to the signal at the first revival.  The revival signal as a function of lattice depth is shown in Fig.~\ref{fig:revival_vs_latticedepth}, where the superfluid to Mott insulator transition is expected to occur around $13E_{\rm R}$~\cite{greiner02}.  In a non-interacting system without any dissipation, one would expect complete revivals.  The decrease in revival fraction as the lattice depth increases is consistent with the picture that as interactions increase, phase fluctuations among neighboring lattice sites increase and consequently the superfluid fraction decreases.  In the Mott insulating phase, revivals should completely disappear, except for a very weak signal at intermediate lattice depths due to particle-hole correlations~\cite{gerbier05}.

To understand the measured revival decay, we have numerically studied the one-dimensional Gross-Pitaevskii equation that assumes interacting matter waves at zero temperature.  The simulations show that interactions and finite size effects have negligible effect on the decay of revivals.  Empirically, randomizing the phase between neighboring sites reduces the revival fraction.  The loss of phase coherence across lattice sites could be due to factors such as temperature, beyond Gross-Pitaevskii equation effects, or technical factors.  We have considered quantum depletion~\cite{xu06}, but the calculated depletion fraction at $5E_{\rm R}$ is 2\%, too low to account for the observed decay.
Future work should provide a more complete picture of the decay of revivals.

Note that Bragg scattered revivals are complementary to the observation of diffraction peaks in time-of-flight absorption images~\cite{greiner02}.  Both are based on matter-wave interference due to superfluid coherence: revivals at short expansion times and diffraction at long expansion times.  Diffraction is a far-field effect only possible for finite size samples, whereas revivals are a near-field bulk effect.  The order of the revival or the angular resolution of the diffraction peaks determine whether these techniques probe short-range or long-range spatial coherence.  Further studies are needed to determine which technique is more sensitive to certain aspects of the superfluid to Mott insulator transition.

In this Letter we have not focused on the effects of occupation number on Bragg scattering.  In principle the atomic wavefunctions are more extended for higher occupation numbers, but this effect is small for our parameters.  However, light scattering at higher occupation numbers will have an inelastic component due to colliding atoms or photoassociation~\cite{gallagher89}.  The dependence of Bragg scattering on lattice depth suggests that these effects are not dominant.

After Bragg scattering has been established as a probe for Mott insulators, it can now be applied to study other types of quantum phases such as antiferromagnetic ordering in both the occupation number and spin sectors~\cite{corcovilos10,simon11}.
Although temperatures required to realize spin ordering are on the order of tens to hundreds of picokelvins, recent experimental results suggest a way forward to lower the temperature of a two-component Mott insulator~\cite{medley11}.

In conclusion, we have observed Bragg scattering of near-resonant photons from a quantum degenerate Bose gas in a three-dimensional optical lattice.  We have shown that this technique probes not only the periodic structure of the atoms, but also the spatial and momentum width of the localized atomic wavefunctions and the superfluid coherence.

We acknowledge Ian Spielman, Eugene Demler, Igor Mekhov, Jesse Amato-Grill, Niklas Jepsen and Ivana Dimitrova for fruitful discussions and technical support.  We acknowledge Yichao Yu for experimental assistance. We thank Jonathon Gillen for a critical reading of the manuscript.  H.M. acknowledges support from the NDSEG program.  This work was supported by the NSF through the Center of Ultracold Atoms, by NSF award PHY-0969731, through an AFOSR MURI program, and under ARO Grant No.W911NF-07-1-0493 with funds from the DARPA OLE program.

\end{document}